\documentclass[floatfix,aps,prb,twocolumn,10pt,superscriptaddress]{revtex4-2}

\usepackage{graphicx}
\usepackage{amsmath}
\usepackage{amssymb}
\usepackage[dvipsnames]{xcolor}
\definecolor{p-channel}{HTML}{78BCD8}
\definecolor{c-channel}{HTML}{D36146}
\definecolor{d-channel}{HTML}{66458C}
\usepackage{mathtools}
\usepackage{hyperref}
\usepackage{float}
\hypersetup{
  colorlinks,
  linkcolor=c-channel!60!black,
  citecolor=d-channel!80!black,
  urlcolor=p-channel!60!black,
}
\usepackage[capitalise]{cleveref}
\usepackage{dsfont}
\usepackage{bm}
\usepackage{makerobust}
\usepackage[mode=buildnew]{standalone}
\usepackage[misc]{ifsym}
\usepackage{academicons}
\definecolor{orcidlogocol}{HTML}{A6CE39}

\newcommand{\bvec}[1]{\boldsymbol{#1}}

\renewcommand{\Im}{\mathrm{Im}}

\newcommand{\approptoinn}[2]{\mathrel{\vcenter{
  \offinterlineskip\halign{\hfil$##$\cr
    #1\propto\cr\noalign{\kern2pt}#1\sim\cr\noalign{\kern-2pt}}}}}
\newcommand{\appropto}{\mathpalette\approptoinn\relax}

\newcommand{\makeauthor}[2]{\newcommand{#1}[1]{{%
  \protect%
  \sffamily\color{#2}{%
    \bfseries\begingroup\escapechar=-1\edef\x{\endgroup\string#1}\x:%
  } ##1}}%
  \MakeRobustCommand#1}
\makeauthor{\LK}{red}
\makeauthor{\JH}{blue}
\newcommand{\dd}{\mathrm{d}}
\newcommand{\vpp}{{\vphantom{\prime}}}
\newcommand{\proj}[2]{#1\big[#2\big]}
\newcommand\bzint[1]{\mathchoice{%
  \int_{\mathrm{BZ}}\!\!\dd\bvec{#1}
  }{%
  \int_{\mathrm{BZ}}\!\dd\bvec{#1}
  }{%
  \int_{\mathrm{BZ}}\!\dd\bvec{#1}
  }{%
  \int_{\mathrm{BZ}}\!\dd\bvec{#1}
  }%
}
\newcommand\bzintA[1]{\mathchoice{%
  \int_{\mathrm{BZ}}\!\!\dd\bvec#1
  }{%
  \int_{\mathrm{BZ}}\!\dd\bvec#1
  }{%
  \int_{\mathrm{BZ}}\!\dd\bvec#1
  }{%
  \int_{\mathrm{BZ}}\!\dd\bvec#1
  }%
}

\newcommand{\orcidid}[1]{%
  \hypersetup{urlcolor=orcidlogocol}%
  \raisebox{0.1em}{\resizebox{!}{0.7\height}{\href{#1}{\color{orcidlogocol}\aiOrcid}}}%
  \hypersetup{urlcolor=RoyalPurple}%
}

\newcommand{\removed}[1]{}
\newcommand{\added}[1]{#1}

\begin{document}

\title{Surface Functional Renormalization Group for Layered Quantum Materials}
\author{Lennart Klebl}
\thanks{\Letter{}~\href{mailto:lennart.klebl@uni-wuerzburg.de}{lennart.klebl@uni-wuerzburg.de}}
\affiliation{Institute for Theoretical Physics and Astrophysics, Universität
Würzburg, Am Hubland, 97074 Würzburg, Germany}
\author{Dante M.~Kennes}
\affiliation{Institut f\"ur Theorie der Statistischen Physik, RWTH Aachen
University and JARA-Fundamentals of Future Information Technology, 52056 Aachen,
Germany}
\affiliation{Max Planck Institute for the Structure and Dynamics of Matter,
Center for Free Electron Laser Science, 22761 Hamburg, Germany}
\date{\today}

\begin{abstract}
  We present an extension to the two-dimensional functional renormalization
  group to efficiently treat interactions on the surface or at interfaces of
  three-dimensional systems.
  As an application, we consider a semi-infinite stack of two-dimensional square lattices, including a Hubbard interaction on the surface layer and an alternating interlayer coupling. We investigate how strongly correlated states of the
  decoupled two-dimensional Hubbard model on the surface evolve under inclusion of such an SSH-like interlayer coupling. For large parts of the
  phase diagram as a function of the interlayer hopping parameters, the
 physics of the  two-dimensional system prevails, with antiferromagnetic, superconducting
  $d$-wave, and ferromagnetic correlations taking center stage. However, for intermediate interlayer couplings the
  superconducting state at intermediate interaction strengths separates into two regimes  by a small   region of incommensurate spin-density-wave
  and spin-bond order, enabling the potential realization of chiral spin-bond order.
\end{abstract}

\maketitle

\section{Introduction}
One of the most studied models of strongly correlated electrons in condensed
matter physics is the two-dimensional Hubbard model~\cite{hubbard1963electron}.
Although being an extremely simplified model, it is believed to capture the
essence of the pairing mechanism relevant for high-$T_\mathrm{c}$ cuprate
superconductors~\cite{anderson1987resonating, emery1987theory,
arovas2022hubbard}. But, even on the level of model physics, obtaining an
unbiased characterization of the different regimes at play remains a difficult
challenge. Among many other highly developed
methods~\cite{qin2020absence, schafer2021tracking, arovas2022hubbard,
qin2022hubbard}, one promising contender at small to intermediate interaction strengths is the functional renormalization group
(FRG)~\cite{wetterich1993exact, morris1994exact}. Formulated as a diagrammatic
technique taking all interchannel feedback into account while successively
integrating out the high-energy degrees of freedom, it provides robust and
unbiased characterization of ordering tendencies~\cite{salmhofer2001fermionic,
metzner2012functional, wang2012functional, platt2013functional,
kennes2018strong, classen2019competing, dupuis2021nonperturbative}.

\begin{figure}
  \centering
  \includegraphics[width=\columnwidth]{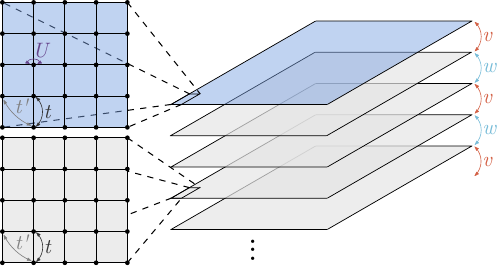}
  \caption{Three-dimensional Hubbard-SSH model. Each of the layers has nearest-
  and next-nearest-neighbor hopping terms $t$ and $t'$, respectively (black and
  gray). The coupling of subsequent layers is given by two alternating hopping
  amplitudes $v$ (orange) and $w$ (light blue), as in an SSH chain. We treat the
  renormalization of the two-particle interaction only in the outermost layer,
  where we include an on-site Hubbard-$U$ term (blue).}
  \label{fig:sketch}
\end{figure}

While studies of multi-orbital or spin-orbit coupled models with FRG have been
extensively carried out~\cite{platt2013functional, kiesel2013unconventional,
delapena2018competing_PHD, klett2021from, klebl2022moire, qin2022functional,
klebl2022competition, beyer2023rashba, he2023superconductivity, profe2024kagome,
profe2024magic, fischer2025theory, beck2026kekule, durrnagel2025altermagnetic,
guo2025angle}, performing FRG simulations even in simple, periodic
three-dimensional systems remains a numerical
challenge~\cite{ehrlich2020functional, profe2024competition,
bigi2024pomeranchuk}. Fortunately, electron-electron interactions are often dominant
within a single layer~\cite{lee2007high} at interfaces or on the surface in the case of
topological three-dimensional systems~\cite{rachel2018interacting} or intralayer physics is only weakly perturbed by interlayer coulings, such as in some van der Waals heterostructures. Therefore,
it is a promising simplification to study how electron-electron interactions
within a single two-dimensional layer (or surface) can induce strongly correlated
phases when the kinetic part of the Hamiltonian is three-dimensional.

Here, we develop a variant of momentum space FRG that treats the renormalization
of interactions on a surface or a single layer, embedded in a three dimensional
system. We call this variant `surface FRG' in the follwing.
\added{This method efficiently captures intralayer interaction
effects in an unbiased manner, allowing to study the consequences of
three-dimensional embedding of quasi two-dimensional systems with significantly
reduced numerical effort.}
We apply \removed{this method}\added{the `surface FRG'{}} to a three-dimensional
Su-Schrieffer-Heeger (SSH) model~\cite{su1979solitons}: \Cref{fig:sketch}
illustrates the stack of square lattice layers coupled with alternating hoppings
in out-of-plane direction. The surface layer (blue) exhibits an on-site
Hubbard-$U$, while all other layers of the semi-infinite ``Hubbard-SSH''{} stack
are  non-interacting. Using surface FRG, we analyze the effects of
electron-electron interactions in the surface layer.

The rest of the paper is structured as follows: We first give an overview of the
surface FRG method in \cref{sec:method}. Thereafter, we introduce the
three-dimensional Hubbard-SSH model in greater detail (see \cref{sec:model}).
The results of our surface FRG calculations are presented in \cref{sec:results}.

\section{Functional renormalization group}
\label{sec:method}
What most FRG schemes of the two- and four-point vertices have in common is that
they require only the vertices at the starting scale $\Lambda_0=\infty$ and the
non-interacting two-point Green's function at renormalization scale $\Lambda$ as
input~\cite{honerkamp2001temperature, metzner2012functional}. Surface Green's
functions that include information on the semi-infinite nature of the system can
be recursively generated from the bulk couplings~\cite{kalkstein1971green,
sancho1985highly}. Consequently, investigations on strong correlations arising
from surface interactions can be carried out efficiently employing FRG with
surface Green's functions as the non-interacting input.

\begin{figure}
  \centering
  \includegraphics[width=\columnwidth]{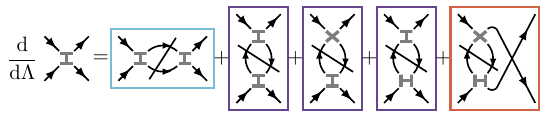}
  \caption{Diagrammatic representation of the flow equation of the four-point
  vertex. The diagrams are grouped into channels corresponding to distinct
  transfer momenta, with the particle-particle channel $\color{p-channel}P^\Lambda$
  in light blue, the direct
  particle-hole channel ${\color{d-channel}D^\Lambda}$ in purple, and the crossed particle-hole channel ${\color{c-channel}C^\Lambda}$ in orange.
  Slashed propagator pairs denote a scale derivative ($\dd/\dd\Lambda$) of the
  corresponding propagator pair. The vertices conserve electron spin along the
  gray lines connecting two in-/out-going electron lines.}
  \label{fig:diagrams}
\end{figure}

As keeping track of all variables is numerically difficult to achieve, many
approximations to and truncations of the FRG equations have been
proposed~\cite{husemann2009efficient, wang2012functional, platt2013functional,
lichtenstein2017high-performance, honerkamp2018efficient,
honerkamp2018limitations, eckhardt2020truncated, hille2020quantitative,
klebl2020functional, hauck2021electronic, profe2022tu2frg}. One of these
truncations---commonly applied for electrons in two dimensional systems---is
keeping vertices only up to the four-point one, i.e., the two-particle
interaction. Additional approximations include going to the static limit
(neglecting frequency dependencies) of the vertices (self-energy and
interaction) as well as disregarding self-energy
feedback~\cite{honerkamp2001breakdown, honerkamp2001electron,
honerkamp2001temperature, platt2013functional}. In order to keep the numerical
effort of our investigations in the present system tangible, we hence %reduce the
%number of degrees of freedom and 
focus on the renormalization of the static
four-point vertex (interactions), while discarding all higher vertices and the
two-point vertex (self-energies). We further make use of spin rotational
invariance, i.e., $SU(2)$ symmetry and assume translational invariance in the
in-plane directions. The effects of out-of-plane hopping are encoded in the
propagators $G_0$ by including an exact self-energy contribution  $\Sigma_0$, such that the resulting flow equation of the four-point vertex
retains the original diagrammatic structure (see \cref{fig:diagrams}):%
\begin{widetext}%
\begin{align}
  \label{eq:frg-equation}
  \frac{\dd V^\Lambda}{\dd \Lambda} &{}= \frac{\dd}{\dd\Lambda}
    \Big( \proj{P^{-1}}{{\color{p-channel}P^\Lambda}} +
          \proj{C^{-1}}{{\color{c-channel}C^\Lambda}} +
          \proj{D^{-1}}{{\color{d-channel}D^\Lambda}} \Big) \,, \\
  \frac{\dd P^\Lambda(\bvec q_P^\vpp,\bvec k_P^\vpp,\bvec k_P^\prime)}
      {\dd \Lambda} &{}= \bzint{k} \,
    \proj{P}{V^\Lambda}(\bvec q_P^\vpp,\bvec k_P^\vpp,\bvec k)
    \dot{L}_-^\Lambda(\bvec q_P,\bvec k)
    \proj{P}{V^\Lambda}(\bvec q_P^\vpp,\bvec k,\bvec k_P^\prime) \,, \\
  \frac{\dd C^\Lambda(\bvec q_C^\vpp,\bvec k_C^\vpp,\bvec k_C^\prime)}
      {\dd \Lambda} &{}= \bzint{k} \,
    \proj{C}{V^\Lambda}(\bvec q_C^\vpp,\bvec k_C^\vpp,\bvec k)
    \dot{L}_+^\Lambda(\bvec q_C,\bvec k)
    \proj{C}{V^\Lambda}(\bvec q_C^\vpp,\bvec k,\bvec k_C^\prime) \,, \\
  \frac{\dd D^\Lambda(\bvec q_D^\vpp,\bvec k_D^\vpp,\bvec k_D^\prime)}
      {\dd \Lambda} &{}= \bzint{k} \,
    \Big\{
      -2\proj{D}{V^\Lambda}(\bvec q_D^\vpp,\bvec k_D^\vpp,\bvec k)
      \dot{L}_+^\Lambda(\bvec q_D,\bvec k)
      \nonumber
      \proj{D}{V^\Lambda}(\bvec q_D^\vpp,\bvec k,\bvec k_D^\prime) + {} \\
      %                                   | alignment spacer
      \proj{D}{V^\Lambda}(\bvec q_D^\vpp, & \bvec k_D^\vpp,\bvec k)
      \dot{L}_+^\Lambda(\bvec q_D,\bvec k)
      \proj{C}{V^\Lambda}(\bvec q_D^\vpp,\bvec k,\bvec k_D^\prime) + 
      \proj{C}{V^\Lambda}(\bvec q_D^\vpp,\bvec k_D^\vpp,\bvec k)
      \dot{L}_+^\Lambda(\bvec q_D,\bvec k)
      \proj{D}{V^\Lambda}(\bvec q_D^\vpp,\bvec k,\bvec k_D^\prime) \Big\} \,.
\end{align}
\end{widetext}%
Here, $V^\Lambda(\bvec k_1,\bvec k_2,\bvec k_3)$ denotes the $SU(2)$ symmetrized
four-point vertex function at scale $\Lambda$ with $\bvec k_1$, $\bvec k_2$ the
two dimensional in-plane momenta of incoming electron lines and $\bvec k_3$, $\bvec k_4=\bvec k_1+\bvec
k_2-\bvec k_3$ the momenta of outgoing electrons, with spin being conserved
along $\bvec k_1 \rightarrow \bvec k_3$ and $\bvec k_2 \rightarrow \bvec
k_4$\removed{.}\added{; in order to describe both direct and exchange processes
we treat the vertex component $V_{\uparrow\downarrow\downarrow\uparrow}$ as
representative, which results in the five diagrams
depicted in \cref{fig:diagrams}~\cite{salmhofer2001fermionic}.}
The integral over the Brillouin zone is assumed to be normalized, $\bzint{k}=1$.
Note that we start the flow at $\Lambda=\infty$, where the system is exactly
solvable due to suppression of fluctuations, and solve the differential equation
for $\Lambda\rightarrow0$, where the full, interacting description is recovered.
As an ordered phase, within the applied truncation scheme, is indicated by a
divergence in the four-point vertex at finite $\Lambda_\mathrm{C}>0$, we stop
the flow if the maximum element of the vertex surpasses a critical value.
Information on the ordered phase is obtained from the vertex at the critical
scale $\Lambda_\mathrm{C}$.

We further define the three channel-projections of the four-point vertex
$\proj{X}{V^\Lambda}$ for each of the channels (particle-particle $P$, crossed
particle-hole $C$, and direct particle-hole $D$) as follows:
\begin{align}
  \proj{P}{V^\Lambda}(\bvec q_P^\vpp,\bvec k_P^\vpp,\bvec k_P^\prime) &{} =
    V^\Lambda(\bvec k_P^\vpp, \bvec q_P^\vpp - \bvec k_P^\vpp, \bvec k_P^\prime)
    \,,\\
  \proj{C}{V^\Lambda}(\bvec q_C^\vpp,\bvec k_C^\vpp,\bvec k_C^\prime) &{} =
    V^\Lambda(\bvec k_C^\vpp, \bvec k_C^\prime - \bvec q_C^\vpp,
    \bvec k_C^\prime) \,,\\
  \proj{D}{V^\Lambda}(\bvec q_D^\vpp,\bvec k_D^\vpp,\bvec k_D^\prime) &{} =
    V^\Lambda(\bvec k_D^\vpp, \bvec k_D^\prime - \bvec q_D^\vpp, \bvec k_D^\vpp
    - \bvec q_D^\vpp) \,,
\end{align}
with $\bvec q_X$ the leading bosonic transfer momentum (Mandelstam variable)
corresponding to channel $X$. The inverse projection can be defined indirectly
via
\begin{equation}
  \proj{X^{-1}}{\proj{X}{V^\Lambda}}(\bvec k_1,\bvec k_2,\bvec k_3) =
  V^\Lambda(\bvec k_1,\bvec k_2,\bvec k_3)
\end{equation}
for $X\in\{P,C,D\}$. The fermionic loop integrals depend on the regulator
chosen. We here assume the sharp frequency cutoff~\cite{beyer2022reference}, for which the
following holds:
\begin{multline}
  \dot L_\pm^\Lambda(\bvec q,\bvec k) = \frac1{2\pi}\big[
    G_0(\pm i\Lambda,\bvec k)G_0( i\Lambda,\pm \bvec k \mp \bvec q) + \\
    G_0(\mp i\Lambda,\bvec k)G_0(-i\Lambda,\pm \bvec k \mp \bvec q)
    \big] \,,
\end{multline}
with $G_0(i\omega,\bvec k)$ the free fermionic Green's function at Matsubara
frequency $i\omega$ and (in-plane) momentum $\bvec k$.

We stress that when treating surface systems, the structure of
$G_0(i\omega,\bvec k)$ is highly non-trivial as it includes the effects of
out-of-plane hopping as frequency-dependent self-energy $\Sigma_0(i\omega, \bvec
k)$~\cite{Karrasch2010functional}. Therefore, in this case, the Matsubara
Green's function no longer satisfies its regular functional form and instead
$G_0(i\omega,\bvec k)^{-1} = i\omega-\epsilon(\bvec k)-\Sigma_0(i\omega,\bvec
k)+\mu$ holds. For the evaluation of the loop integrals $\dot L^\Lambda_\pm$, it
therefore is crucial to use the sharp frequency cutoff---so we only need to
evaluate $G_0(\pm i\Lambda,\bvec k)$ for a single $\Lambda$ during each step of
the solution of \cref{eq:frg-equation}.

\section{The 3D Hubbard-SSH model}
\label{sec:model}
We demonstrate how FRG can be applied to surface interactions by constructing a
three-dimensional Hubbard-SSH model. \Cref{fig:sketch} visualizes the setup:
Within each two-dimensional layer $l$ of the semi-infinite stack, we employ a
next-nearest-neighbor tight-binding model on the square lattice with a (kinetic)
Hamiltonian
\begin{equation}
  H_\mathrm{2D}^l = \sum_{ij,\sigma} \,t_{ij}\,c^\dagger_{l,i,\sigma}
  c^{\vphantom{\dagger}}_{l,j,\sigma} \,,
\end{equation}
where the hopping amplitudes are given by $t_{ij} = t$ for nearest-neighbor
sites and $t_{ij} = t'$ for next-nearest-neighbor sites. The operator
$c^{(\dagger)}_{l,i,\sigma}$ annihilates (creates) an electron with spin
$\sigma$ at site $i$ in layer $l$. We couple adjacent sites in subsequent layers
as an SSH-chain~\cite{su1979solitons} with alternating hopping amplitudes $v$
and $w$. As we assume the stack to be semi-infinite, the layer index ranges from
$l=0$ to $l=\infty$ with $l=0$ denoting the surface layer. We further choose the
coupling from the surface to the first bulk layer to be $t_0=v$. The full
non-interacting Hamiltonian thus reads
\begin{equation}
  H_0 = \sum_{l=0}^\infty \bigg[ H^l_\mathrm{2D} + \sum_{i,\sigma}
  t_l \big( c^\dagger_{l,i,\sigma} c^{\vphantom{\dagger}}_{l+1,i,\sigma} +
  \mathrm{h.c.} \big) \bigg] \,,
\end{equation}
with $t_l=v$ for $l$ even and $t_l=w$ for $l$ odd. We can transform the
two-dimensional dependency of $H_0$ to momentum space and write the coupling of
subsequent layers as a semi-infinite matrix to arrive at the following instructive
representation of $H_0$:
\begin{widetext}
\begin{equation}
  H_0 = \sum_{\bvec k,\sigma}
  \begin{pmatrix}
    c^\dagger_{0,\bvec k,\sigma} &
    c^\dagger_{1,\bvec k,\sigma} &
    \cdots
  \end{pmatrix}
  \cdot
  \hat H_0(\bvec k)
  \cdot
  \begin{pmatrix}
    c^{\vphantom{\dagger}}_{0,\bvec k,\sigma} \\
    c^{\vphantom{\dagger}}_{1,\bvec k,\sigma} \\
    \vdots
  \end{pmatrix} = \sum_{\bvec k,\sigma}
  \begin{pmatrix}
    c^\dagger_{0,\bvec k,\sigma} &
    c^\dagger_{1,\bvec k,\sigma} &
    \cdots
  \end{pmatrix}
  \cdot
  \begin{pmatrix}
    h(\bvec k) & v \\
    v          & h(\bvec k) & w \\
               & w          & h(\bvec k) \\
               &            &           & \ddots \\
  \end{pmatrix}
  \cdot
  \begin{pmatrix}
    c^{\vphantom{\dagger}}_{0,\bvec k,\sigma} \\
    c^{\vphantom{\dagger}}_{1,\bvec k,\sigma} \\
    \vdots
  \end{pmatrix} \,,
\end{equation}
\end{widetext}
where we defined the in-plane dispersion
\begin{equation}
  h(\bvec k) = -t\big(\cos(k_x)+\cos(k_y)\big) - 4t'\cos(k_x)\cos(k_y) \,.
\end{equation}
For our simulations, we fix the chemical potential to the Van Hove singularity
of a single layer, i.e., $\mu=-4t'$ and further set the energy unit as $t=1$.
As indicated in \cref{fig:sketch}, we add a Hubbard-interaction to the surface
layer. The interacting part of the Hamiltonian therefore has nonzero components
only for $l=0$ and reads:
\begin{equation}
  H_U = U\,\sum_{i} n_{0,i,\uparrow}n_{0,i,\downarrow} \,.
\end{equation}

The renormalization procedure of the surface interactions requires the free
surface Green's function at each step of the iteration. We can employ the scheme
from Ref.~\cite{sancho1985highly} (implemented, e.g., in
Ref.~\cite{wu2017wanniertools}) to obtain the surface Green's function
$G_0(\omega,\bvec k)$ for both real and imaginary frequencies~\footnote{
  In fact, the implementation provided in Ref.~\cite{wu2017wanniertools}
  not only provides surface Green's functions, but at the same time local bulk
  Green's functions. The bulk Green's function can be used in the same manner as
  the surface Green's function without further modifications to the FRG
  equations to study intralayer interactions.
}. To use the highly convergent algorithm, we must bring $H_0$ in a form of
periodically repeating couplings $\hat{\mathcal H}$ on the diagonal and
(repeating) tunneling matrices $\hat{\mathcal T}$ between these diagonal
couplings. Therefore we define
\begin{align}
  \hat{\mathcal H}(\bvec k) &{}= \begin{pmatrix} h(\bvec k) & v \\ v & h(\bvec
  k) \end{pmatrix} \,,\\
  \hat{\mathcal T} &{}= \begin{pmatrix} 0 & 0 \\ w & 0 \end{pmatrix}
\end{align}
and re-introduce the required periodic tridiagonal structure of the
tight-binding matrix $\hat H_0(\bvec k)$:
\begin{equation}
  \hat H_0(\bvec k) = \begin{pmatrix}
    \hat{\mathcal H}(\bvec k) & \hat{\mathcal T} & \\
    \hat{\mathcal T}^\dagger & \hat{\mathcal H}(\bvec k) & \hat{\mathcal T} \\
    & \hat{\mathcal T}^\dagger & \hat{\mathcal H}(\bvec k) & \\
    &&& \ddots
  \end{pmatrix} \,.
\end{equation}
The such obtained surface Green's function $\hat{\mathcal G}_0(\omega, \bvec k)$
will then be a $2\times2$ matrix. The $(1,1)$ (top left) component yields the
desired Green's function of the top layer: $G_0(\omega,\bvec k) = \mathcal
G^{1,1}_0(\omega,\bvec k)$.

\begin{figure}
  \includegraphics{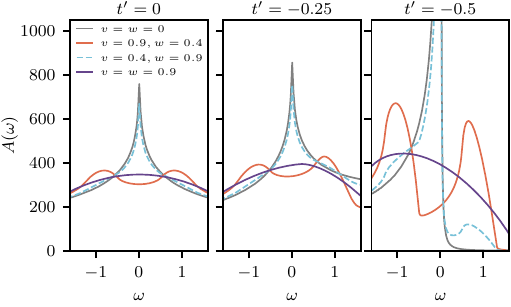}
  \centering
  \caption{Non-interacting spectral density for three cases of
  $t'\in\{0,-0.25,-0.5\}$. The chemical potential is fixed at $\mu=-4t'$ such
  that the system is at Van Hove filling in the layer-decoupled limit ($\omega$ is measured with respect to $\mu$). For each
  panel, the four lines represent four different choices of the out-of-plane
  hopping parameters $v$ and $w$: The gray line corresponds to the
  layer-decoupled system ($v=w=0$), the orange line to $v=0.9$ and $w=0.4$, the
  dashed cyan line to $v=0.4$ and $w=0.9$, and the purple line to an almost
  isotropic three-dimensional system with $v=w=0.9$.}
  \label{fig:densities}
\end{figure}

To characterize the non-interacting system, we consider the non-interacting
surface spectral function of the 3D Hubbard-SSH model before analyzing the effect of a
surface Hubbard-$U$ interaction. The spectral function is given by
\begin{equation}
  A(\omega) = \removed{-}\frac1\pi\,\Im \bzint{k}\,G_0(\omega\removed{-}\added{+}i\eta,\bvec k) \,,
\end{equation}
where we introduced a broadening parameter $\eta$ that we set to $10^{-2}$.
\Cref{fig:densities} showcases spectral functions for three values of $t'$ and
four combinations of $(v,w) \in \{(0,0), (0.9,0.4), (0.4,0.9), (0.9,0.9)\}$. At
$v=w=0$ the system is effectively two-dimensional leading to the characteristic
Van Hove singularity. As long as a surface state is present, i.e., the system is
 topological where $w>v$, the Van Hove singularity remains at
$\omega=0$ (we measure $\omega$ with respect to the chemical potential $\mu$). For the other case, $v\geq w$, the single peak is split into two
peaks and smeared out due to the lack of a surface state. Thus we expect the
tendency towards correlations driven by electron-electron interactions to be
reduced for $v\geq w > 0$.

\section{Results}
\label{sec:results}
\begin{figure}
  \centering
  \includegraphics{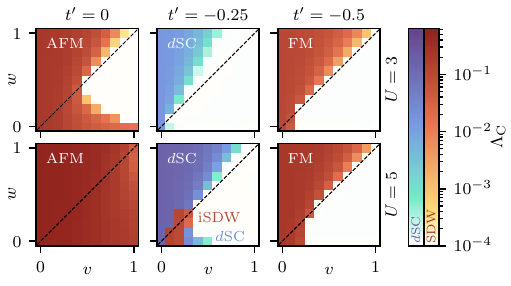}
  \caption{Critical scale $\Lambda_\mathrm{C}$ (approximately onset temperature
  of the corresponding order) and type of leading instability for
  $t'\in\{0,-0.25,-0.5\}$ and $U\in\{3,5\}$. The out-of-plane hopping amplitudes
  $v$ and $w$ are varied from zero to one on a regular $11\times11$ grid. The
  red color-map corresponds to spin-density-wave order (SDW)\added{, which
  includes antiferromagnetism (AFM), incommensurate
  spin-density-waves/spin-bond-order
  (iSDW), and ferromagnetism (FM);} and the blue
  color-map to \added{$d$-wave} superconducting order (\added{$d$}SC).
  Literature results for the two-dimensional Hubbard model are reproduced for
  $v=0$, where the surface is decoupled from the bulk.
  }
  \label{fig:phases}
\end{figure}
We carry out FRG simulations for $t'\in\{0,-0.25,-0.5\}$ on a $11\times11$ grid
for $v,w\in[0,1]$. Moreover, we investigate the two
cases $U=3$ and $U=5$. We integrate the FRG flow [\cref{eq:frg-equation}] using
an adaptive Euler scheme and calculate the vertex function on a $20\times20$
Bravais momentum mesh, refined with $45\times45$ points in the loop
integrals~\footnote{%
  For details on the numerical methods see Ref.~\cite{beyer2022reference}. The
  open-source implementation of the Hubbard-SSH model FRG~\cite{code} makes use
  of the publicly available divERGe library~\cite{profe2024diverge,
  profe2024diverge-code}.
}. We consider the flow diverged when the maximum element of the four-point
vertex reaches $3\cdot10^{1}$. The corresponding scale at which the divergence
occurs is labelled by $\Lambda_\mathrm{C}$. With our choice of the sharp cutoff
as regulator, the value of $\Lambda_\mathrm{C}$ roughly corresponds to an onset
temperature of the order. By inspection of which channel causes the divergence,
we can differentiate between spin- and charge-density waves and superconducting
instabilities. Analysis of the four-point vertex at the critical scale enables
us to obtain more details about the instability, i.e., the leading ordering
vector or symmetry of the superconducting state.

\Cref{fig:phases} shows a $v$-$w$ false color plot of the critical scale for different values
of $t'$ and $U$, where the red and blue color-maps indicate whether a
superconducting or a magnetic instability is dominating. The  FRG
result for the two-dimensional Hubbard model, i.e., spin-density wave
(SDW) instabilities at $t'=0$ and $t'=-0.5$ as well as a superconducting (SC)
instability at $t'=-0.25$ is reproduced for the $v=0$ case, where the surface
layer is decoupled from the bulk. Furthermore, we observe that in most regimes of
the phase diagram, the type of instability is independent of $v$ and $w$, up to the point where the coupling to the bulk becomes strong enough to suppress order entirely for $v>w$. Only
for $U=5$ and $t'=-0.25$, a small area of SDW ordering emerges upon setting $v$
and $w$ to small, nonzero values. Moreover, the presumption of a reduced
tendency towards strongly correlated states for $v\geq w$ is mostly fulfilled,
except for the $t'=0$ case, where the SDW instability is strong enough to
prevail in the bilayer case (with $w=0$ and $v=1$). In addition, we note that
in the limit towards a homogeneous three dimensional system surface states have
a weakened spectral density, which leads to decreased ordering tendencies along
the diagonal line in $(v,w)$ phase space. \added{The prevalence of order in the
upper left triangle ($w>v$) of the phase diagrams in \cref{fig:phases} is tied to
the presence of the topological surface state.}

We analyze the SDW phases in greater detail by calculating the interacting
magnetic susceptibility $\chi_f(\bvec q)$ from the vertex at the critical scale
$\Lambda_\mathrm{C}$:
\begin{multline}
  \chi_f(\bvec q) = \bzint{k} f(\bvec k) \bzintA{k'} f(\bvec k') \,
  \chi_0(\bvec q,\bvec k) \\
  \proj{C}{V^{\Lambda_\mathrm{C}}}(\bvec q,\bvec k,\bvec k')
  \chi_0(\bvec q,\bvec k') \,,
\end{multline}
where we used the non-interacting susceptibility
\begin{equation}
  \chi_0(\bvec q,\bvec k) = \frac1\beta \sum_{ik_0} G_0(q+k)G_0(k)\,,
\end{equation}
and set the inverse temperature to $\beta\equiv1/\Lambda_\mathrm{C}$. The
formfactors $f(\bvec k)$ belong to the $C_{4v}$ symmetrized lattice harmonics of
the square lattice~\cite{platt2013functional}, see \cref{tab:formfactors} for an
overview.
\begin{table}%
  \caption{Symmetrized lattice harmonics for the two-dimensional square lattice
  with $C_{4v}$ rotational symmetry. Each row corresponds to a different
  irreducible representation of the formfactor. The columns label the nearest-neighbor shell $i$ needed to generate the formfactor.}
  \label{tab:formfactors}
  \begin{ruledtabular}
    \begin{tabular}{cccc}
      irr.~rep. & $i=0$ & $i=1$ & $i=2$ \\\hline
      $f^{A_1}_i(\bvec k)$ & 1 & $\cos(k_x)+\cos(k_y)$ & $\cos(k_x)\cos(k_y)$ \\
      $f^{A_2}_i(\bvec k)$ & - & - & - \\
      $f^{B_1}_i(\bvec k)$ & - & $\cos(k_x)-\cos(k_y)$ & - \\
      $f^{B_2}_i(\bvec k)$ & - & - & $\sin(k_x)\sin(k_y)$ \\
      $f^{E_1}_i(\bvec k)$ & - & $\sin(k_x)$ & $\sin(2k_x)$ \\
      $f^{E_2}_i(\bvec k)$ & - & $\sin(k_y)$ & $\sin(2k_y)$ \\
    \end{tabular}
  \end{ruledtabular}
\end{table} %
For $t'=0$ the interacting susceptibility $\chi_f(\bvec q)$ is strongly
peaked at $\bvec q=(\pi,\pi)$ for the on-site and next-nearest-neighbor
formfactors $f = f_0^{A_1}, f_2^{A_1}$, indicating antiferromagnetic order. At
$t'=-0.5$, the dominant contribution originates from $\bvec q=(0,0)$ pointing
towards ferromagnetic order. These two observations are independent on
the out-of-plane couplings $v$ and $w$ (with the points considered being those
ones where the flow diverges to strong coupling).

\begin{figure}
  \centering
  \includegraphics{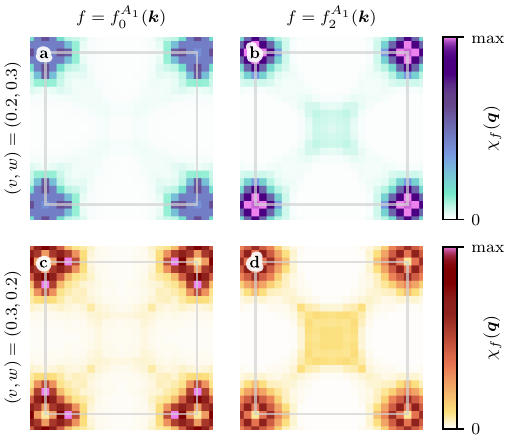}
  \caption{Magnetic susceptibilities $\chi_f(\bvec q)$ generated from the vertex
  at the end of the flow. The Brillouin zone is indicated as a square, with
  incommensurate peaks (magenta) close to $\bvec q=(\pi,\pi)$ for all cases.
  In the momentum contraction of fermion bilinears with the vertex, both a
  constant~(a,c) and a nontrivial~(b,d) formfactor are used. For the topological
  case $(v,w)=(0.2,0.3)$ in~(a,b), the susceptibility is slightly stronger
  for the nontrivial formfactor~(b). For the topologically trivial case
  $(v,w)=(0.3,0.2)$ in~(c,d) the opposite applies and the susceptibility
  is slightly stronger for the constant formfactor~(c). Having a dominant
  instability with non-trivial formfactor (cf.~(b)) can lead to chiral
  superpositions of degenerate wave-vectors $\bvec q$ in the resulting order
  parameter.}
  \label{fig:susc}
\end{figure}

For the small area of SDW phases at $t=-0.25$ and $U=5$ (see the red points in
the central lower panel of \cref{fig:phases}\added{, cf.~Appendix~\ref{app:robust} for the
stability of this region's presence}), incommensurate magnetic instabilities are
formed. In \cref{fig:susc}, we show the corresponding magnetic susceptibilities
for the two topologically distinct cases $(v,w)=(0.2,0.3)$ in~(a,b) and
$(v,w)=(0.3,0.2)$ in~(c,d). In panels~(a,c), we show the constant formfactor
($f=f_0^{A_1}$) contribution, whereas panels~(b,d) depict the contribution of
the second-nearest-neighbor formfactor $f=f_2^{A_1}$. Interestingly, we observe
a transition of relative weights from the non-trivial formfactor in the
topological case~(b) to the constant formfactor in the non-topological case~(c).
The incommensurate nature of the instabilities allows for linear combination of
degenerate momenta in the order parameters of a subsequent mean-field
decoupling. Combining these incommensurate momenta with non-trivial formfactors
in the fermion bilinears as in~(b) may lead to a minimization of the free energy
by chiral superpositions of order parameters, i.e., a coplanar (instead of
collinear) magnetic state.

Superconductivity is observed at $t'=-0.25$ (as expected from the monolayer
limit). We find a $d$-wave symmetric solution $\Delta(\bvec k)\appropto
f_1^{B_1}(\bvec k)$ of the linearized gap equation (see
Refs.~\cite{beyer2022reference, klebl2022moire, klebl2022competition}) for
all SC instabilities. In the case of slightly larger $U=5$, the two
superconducting regions for $w\approx0$ and $v\leq w$ are divided by the SDW
region discussed above. Nevertheless, the leading superconducting instability
does not differ from one region to the other.

\section{Conclusion}
In this manuscript, we presented a novel application of the momentum space
functional renormalization group to study strongly correlated electrons on
surfaces. To this end, we conducted a study of a model system of layered square
lattices coupled in out-of-plane direction with alternating hoppings $v$ and $w$
(``Hubbard-SSH''{} model). While obtaining the well-known two-dimensional limit
for a decoupled surface layer at $v=0$, we explore the phase diagram for $0\leq
v,w\leq1$, interaction strengths $U\in\{3,5\}$, and various
next-nearest-neighbor hoppings $t'$ (all in units of the nearest-neighbor
hopping). We find orders inherited from the monolayer limit: An extended
antiferromagnetic region at $t'=0$, superconducting regions at $t'=-0.25$ and
ferromagnetism at $t'=-0.5$. In the intermediate coupling regime $U=5$, the
superconducting region is separated by a small area of incommensurate
spin-density wave order for weakly coupled layers (i.e., small $v$ and $w$).

Notably, some of these spin density wave instabilities are peaked at
incommensurate transfer momentum $\bvec q$ and at the same time have dominant
contributions in non-trivial formfactor shells. The fermion bilinears
corresponding to the order are
\begin{equation}
  \Delta^\mathrm{SDW}_{\bvec q}(\bvec k) \propto f^{A_1}_2(\bvec k)\,\langle
  c^\dagger_{0,\bvec k+\bvec q,\sigma}c^{\vphantom\dagger}_{0,\bvec k,\sigma'}
  \rangle \,,
\end{equation}
with equal weight on the degenerate values of $\bvec q$. Such incommensurate
spin-density-wave order on the next-nearest-neighbor formfactor $f^{A_1}_2(\bvec
k)$ could potentially lead to complex superpositions, i.e., chiral spin-bond
order. As a next step, we suggest to scrutinize the chiral spin-density-wave
ordering in a mean-field decoupling, potentially combining it with the vertex
obtained from FRG. \added{We note that this poses a numerical challenge, as
(i)~the full three-dimensional structure needs to be considered in order to
obtain the free energy at fixed filling and (ii)~the value of $\bvec q$
close to $(\pi,\pi)$ requires large supercells to render it commensurate.}

Furthermore, the current study could be extended by the following points: First,
we propose to include longer-ranged interactions. These are known to drive
charge-density wave ordering in the two-dimensional case, so we hypothesize that
they could transit to charge-bond ordered phenomena (given the behavior of the
spin-density-wave). Second, it could be instructive to incorporate frequency
dependent self-energies in the flow to analyze the evolution of quasiparticle
weights under interactions. Third, one could include interaction effects of
the bulk on mean-field level, i.e., use $\Sigma_\mathrm{MF}$ instead of
$\Sigma_0$ for the calculation of the surface Green's function.
In the future, we plan to employ the surface FRG to
characterize interaction effects in an unbiased manner in more realistic models,
including the renormalization of interactions in correlated
subspaces~\cite{bigi2024pomeranchuk}. Some intriguing examples include surface
states in Weyl semimetals~\cite{laubach2016density, armitage2018weyl,
xiong2021spin, kundu2022broken, liu2022observation} or (three dimensional)
topological insulators~\cite{neupert2015interacting, lundgren2017nematic,
rachel2018interacting}. The incorporation of phonon-induced interactions into
the FRG would allow to investigate the interplay of
conventional~\cite{maeland2025phonon, maeland2025mechanism} and unconventional
mechanisms for surface superconductivity in systems like PtBi\textsubscript2 and
beyond. Notably, the methodology presented within the scope of this work is
straightforwardly applicable to the study of interface effects in three-dimensional
heterostructures, and as such could shed light on, e.g., superconductivity at the
LAO/STO interface~\cite{gariglio2015electron, scheurer2015topological}.

\begin{acknowledgments}
  We thank J.~B.~Profe for useful discussions. We acknowledge funding by the
  Deutsche Forschungsgemeinschaft (DFG, German Research Foundation) within the
  Priority Program SPP 2244 ``2DMP''{-443274199}, the Würzburg-Dresden Cluster of
  Excellence on Complexity, Topology, and Dynamics in Quantum Matter (ctd.qmat,
  Project ID~390858490, EXC2147), and through the Research Unit QUAST
  (Project ID~449872909, FOR5249). In addition, we acknowledge computational
  resources provided by the Max Planck Computing and Data Facility and through
  JARA on the supercomputer JURECA~\cite{JSC} at Forschungszentrum Jülich.
\end{acknowledgments}

\nocite{code}

\bibliography{ref}

\added{\appendix\onecolumngrid\clearpage
\section{Robustness of the $t'=-0.25$ iSDW}
\label{app:robust}
To clarify that the presence of the iSDW state at $t'=-0.25$ is not a numerical
artifact, we perform truncated unity FRG simulations (via the divERGe
library~\cite{profe2024diverge, profe2024diverge-code}) for various momentum
resolutions, formfactor cutoff distances, and divergence threshold.
\Cref{fig:tufrg} demonstrates that irrespective of momentum resolution, the fact
that these are truncated unity simulations as opposed to the grid FRG results in
the main text, their formfactor cutoff, and their divergence threshold, the iSDW
phase occupies a significant region of phase space.
\begin{figure}[h]
  \includegraphics{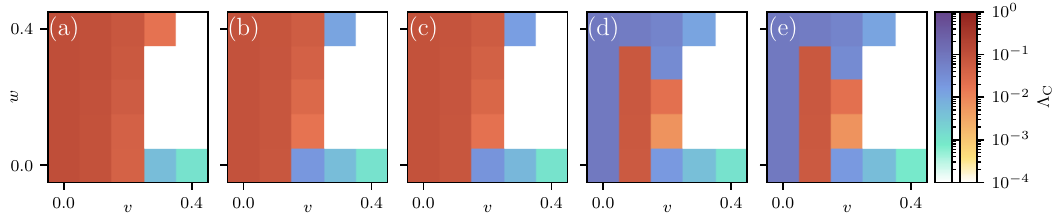}
  \caption{\added{Presence of the iSDW state (red) and the $d$SC state (blue) in
  truncated unity FRG~\cite{profe2022tu2frg} for various parameter settings at
  $t'=-0.25$. For all panels, the coarse momentum grid is fixed at $48\times48$
  points. The fine momentum grid is set to $5\times5$~(a,b,d) and
  $15\times15$~(c,e). Panels~(c,d,e) have the divergence threshold set to $60$,
  while it is $30$/$45$ in~(a)/(b). The formfactor cutoff distance is $2$ in
  all panels except for (c), where it is $1$.}}
  \label{fig:tufrg}
\end{figure}
}

\end{document}